\documentstyle[preprint,tighten,aps]{revtex}

\newcommand{\slsh}[1]{\slash \!\!\! #1}
\begin{document}
\preprint{TRI-PP-99-30, MKPH-T-99-21}
\draft
\title{Field transformations and simple
models illustrating the impossibility
of measuring off-shell effects}
\author{H.\ W.\ Fearing$^1$ and S. Scherer$^2$}
\address{$^1$ TRIUMF, 4004 Wesbrook Mall, Vancouver, British Columbia, 
Canada V6T 2A3}
\address{$^2$ Institut f\"ur Kernphysik, Johannes Gutenberg-Universit\"at,
J.\ J.\ Becher-Weg 45, D-55099 Mainz, Germany}
\date{September 29, 1999}
\maketitle
\begin{abstract}
   In the context of simple models illustrating field transformations
in Lagrangian field theories we discuss the impossibility of measuring
off-shell effects in nucleon-nucleon bremsstrahlung, Compton scattering,
and related processes.
   To that end we introduce a simple phenomenological Lagrangian describing
nucleon-nucleon bremsstrahlung and perform an appropriate change of variables
leading to different off-shell behavior in the nucleon-nucleon amplitude as
well as the photon-nucleon vertex.
   As a result we obtain a class of equivalent Lagrangians, generating
identical $S$-matrix elements, of which the original Lagrangian is but one
representative.
   We make use of this property in order to show that what appears as an
off-shell effect in an $S$-matrix element for one Lagrangian may originate
in a contact term from an equivalent Lagrangian.
   By explicit calculation we demonstrate for the case of nucleon-nucleon
bremsstrahlung as well as nucleon Compton scattering the equivalence of
observables from which we conclude that off-shell effects cannot
in any unambiguous way be extracted from an $S$-matrix element.
   Finally, we also discuss some implications of introducing off-shell
effects on a phenomenological basis, resulting from the requirement
that the description of one process be consistent with that of other
processes described by the same Lagrangian.
\end{abstract}
\pacs{13.40.Gp,13.75.Cs,13.60.Fz,25.40.-h}

\section{Introduction}
\label{sect:intro}

There has been a long history, within the context of calculations in
medium-energy physics, of attempts to find effects of off-shell contributions
in a particular process. Perhaps the prime example of this is nucleon-nucleon
bremsstrahlung which has long been considered the best way to get information
about the off-shell nucleon-nucleon amplitude. In an abstract mathematical
sense, such an off-shell amplitude can be calculated from any potential, say by
solving the Lippman-Schwinger equation, and the hope has always been that one
might be able to distinguish between potentials which are equivalent on shell
via a measurement of their off-shell behavior.

For nucleon-nucleon bremsstrahlung there have been a large number of
calculations, mostly in nonrelativistic potential models which have this aim
\cite{Workman86,Fearing87,Theory,Jetter95,Eden96}. The usual procedure is to
calculate an off-shell nucleon-nucleon amplitude as a separate building block
and attach photons to the external legs of this amplitude. The photon-nucleon
vertex may also have components involving off-shell nucleons. Usually the
so-called double-scattering contribution, involving two strong scatterings with
the photon attached in between, is also included. Less important corrections
such as some relativistic effects, Coulomb corrections, and in some cases
specific exchange-current diagrams are also included.

These state-of-the-art calculations are then compared with experiment. Most
early experiments explored kinematics which were not far enough off shell to
show anything, but the more recent ones, in particular the TRIUMF experiment
\cite{Michaelian90} seemingly showed the need for off-shell effects at least
within the context of current theories. A number of new experiments have been
started in the past few years all designed at least in part to be more
sensitive to kinematics in which the nucleons are as far off shell as possible
\cite{Przewoski,Nomachi,Bilger,Zlomanczuk,Huisman}. These kinematics correspond
to higher photon energies and in general to smaller opening angles between the
two outgoing protons. Thus for example some of the new experiments have been
designed to capture essentially all of the forward-going protons and thus get
opening angles of only a few degrees.

Another case of interest in medium-energy physics where off-shell effects
supposedly enter and have been considered is given by the electromagnetic
interaction of a bound nucleon. Traditionally, electron-nucleus scattering
experiments have been interpreted in terms of the free nucleon current operator
in combination with some recipe to restore gauge invariance.  It has only been
recently that the influence of off-shell effects in the electromagnetic vertex
on an interpretation of (e,e$^\prime$) and (e,e$^\prime$p) data has been
investigated \cite{Naus87,Naus90,Tiemeijer90,Song92}.  Off-shell effects at the
electromagnetic vertex have also been considered for nucleon-nucleon
bremsstrahlung and other processes starting with the early paper of Nyman
\cite{Nyman70} and continuing with the more recent works of Refs.\
\cite{Herrmann,Doenges,Kondratyuk,Li}.

Further processes which might be considered as a source of off-shell
information are two-step processes involving the nucleon such as pion photo-
and electroproduction \cite{Bos92} or real and virtual Compton scattering on
the nucleon \cite{Korchin97}.  Similarly to the bremsstrahlung case the
intermediate nucleon (or pion) in the pole diagrams is off shell and one might
think of exploring the sensitivity of observables to the way the corresponding
vertices behave off shell.

In all of these situations it is common to make some sort of model which
generates off-shell effects in a vertex when that vertex is considered in
isolation. In the nucleon-nucleon bremsstrahlung case almost any potential can
generate a nucleon-nucleon T-matrix which can be extended off shell in some way
determined by the potential. For the photon-nucleon vertex, which would appear
in both bremsstrahlung and in Compton scattering, one can generate off-shell
behavior from a simple phenomenological Lagrangian, or equivalently
parameterize its off-shell structure by some sort of form factor.  In Refs.\
\cite{Naus87,Naus90,Tiemeijer90,Song92,Bos93,Kondratyuk99} the off-shell
effects were generated in terms of more or less sophisticated meson-loop
calculations.  In many other processes off-shell form factors are included. For
example modern nucleon-nucleon potentials put in form factors at the
meson-nucleon vertices, and one question which is at times heatedly discussed
is the appropriate range of such form factors.

Thus within the context of standard calculations of medium-energy processes the
concept of some sort of off-shell effect at the vertices is rather pervasive.

In contrast to this situation, within the context of field theory it has long
been known that there is a certain ambiguity in the evaluation of off-shell
effects. One can for example make field transformations, that is changes of
representation of the fields involved in the Lagrangian, which do not change
any measured quantity, but which can in fact change the off-shell behavior of a
vertex building block of the process
\cite{Haag58,Chisholm61,Kamefuchi61,Coleman69}. One also has the rather simple
observation that, in an amplitude arising from a Feynman diagram, off-shell
behavior at a vertex is cancelled by a similar factor coming from the
intermediate propagator, resulting in an amplitude which could have been
generated by a contact interaction, without reference to off-shell
processes. This was already observed by Gell-Mann and Goldberger in their
derivation of the Compton scattering low-energy theorem \cite{GellMann54}.

It is only relatively recently that these field-theory concepts have begun to
be applied to gain an understanding of the role of, and the ambiguities in, the
off-shell effects which are normally included in medium-energy processes. For
example, in the context of chiral perturbation theory (ChPT), the off-shell
electromagnetic vertex of the nucleon and pion were calculated in Refs.\
\cite{Bos93,Rudy94}, respectively, by including in the Lagrangian certain terms
proportional to the lowest-order equation of motion.  For the pion it was shown
explicitly \cite{Scherer95a} that this off-shell form factor did not contribute
to pion Compton scattering.  Likewise, in a similar model for spin zero
bremsstrahlung it was shown \cite{Fearing98} that off-shell effects arising
from such equation-of-motion terms in the Lagrangian again could be transformed
away, or alternatively replaced by contact terms which did not generate
off-shell contributions at either strong or electromagnetic vertices. Thus in
this context off-shell effects were shown to be unmeasurable, in contrast to
the standard expectation for bremsstrahlung.  Off-shell form factors have been
calculated via dispersion relations as well \cite{Nyman70,Bincer60,BosPhD} and
here it was shown \cite{Davidson96} that the ambiguity in such effects
corresponding to a freedom of choice of field representation was reflected in
an ambiguity in the number of subtraction constants required in the dispersion
relation.

In the present paper we continue and extend in several ways this discussion. In
the following section, Section \ref{sect:zero}, we look at an alternative model
for the spin zero case which illustrates in a simple way the essential features
of the more complicated model of Ref.\ \cite{Fearing98}.

We then extend the approach to look at a simple spin one-half model which is
much more closely allied to the types of phenomenological models which have
been used than that of previous discussions. In particular, it does not involve
ChPT, and so does not depend on any of the formalism there or in particular on
being able to make and truncate an expansion in some small parameter. It
involves spin 1/2 particles, in particular nucleons, as well as photons and
thus should remove any lingering uncertainty that the results of the previous
works somehow depended on the simplicity of a spin zero process. It is more
'realistic' in the sense that it corresponds fairly closely to some
phenomenological models being used to examine off-shell effects. However it is
still sufficiently simple that we can focus on the principles rather than the
details.

In Sec.\ \ref{sect:half} we will discuss the model, which is a somewhat
simplified model Lagrangian for nucleons and photons, and calculate the leading
contributions to nucleon-nucleon bremsstrahlung and Compton scattering on the
nucleon. Section \ref{sect:trans} is devoted to a general discussion of field
transformations and changes of representation and the way they lead to
equivalent Lagrangians. In Sec.\ \ref{sect:newlag} we apply a specific field
transformation to our starting Lagrangian to generate a new Lagrangian which is
closely allied to commonly used phenomenological Lagrangians and which
generates off-shell effects at both strong and electromagnetic
vertices. Section \ref{sect:brem} is devoted to a calculation of bremsstrahlung
and Sec.\ \ref{sect:compton} to a calculation of Compton scattering with the
new Lagrangian.  In both cases we can see explicitly how such off-shell effects
are really not physically measurable quantities. The last section is then
devoted to a summary and some conclusions.

\section{Simple Example}
\label{sect:zero}

   As a first example let us consider the case of $\pi\pi$ bremsstrahlung
which allows us to introduce the main concepts while avoiding complications
due to spin.
   To be specific, we discuss the reaction
$\pi^++\pi^0\to\pi^++\pi^0+\gamma$ in the framework of the nonlinear
$\sigma$ model.
   The present treatment somewhat simplifies results already discussed in
Refs.\ \cite{Scherer95a,Fearing98} in the sense that it will not make use
of higher orders in the momentum expansion of chiral perturbation theory
(ChPT).
   In other words, a discussion only in terms of tree-level diagrams
originating from the nonlinear $\sigma$ model turns out to be sufficient.

   The nonlinear $\sigma$ model Lagrangian, describing pion interactions at
low energies, is given by
\begin{equation}
\label{nlsm}
{\cal L}=\frac{F^2}{4}\mbox{Tr}\left[D_\mu U (D^\mu U)^\dagger\right]
+\frac{F^2 m_\pi^2}{4}\mbox{Tr}(U+U^\dagger),
\end{equation}
where $F=92.4\,\mbox{MeV}$ denotes the pion-decay constant,
$m_\pi=135\, \mbox{MeV}$ is the pion mass,  and the pion
fields are contained in the SU(2) matrix $U$.
   The interaction with the electromagnetic field is generated through
the covariant derivative
$D_\mu U=\partial_\mu U +\frac{i}{2} e A_\mu[\tau_3,U]$, where
$e^2/4\pi \approx 1/137$ and $e>0$.
   At this point we still have a choice how to represent the matrix
$U$ in terms of pion field variables.
   We will make use of two different parameterizations of $U$,
\begin{eqnarray}
\label{u1}
U(x)&=&\frac{1}{F}\left[\sigma(x)+i\vec{\tau}\cdot\vec{\pi}(x)
\right],\quad \sigma(x)=\sqrt{F^2-\vec{\pi}^2(x)},
\\
\label{u2}
U(x)&=&\exp\left[i\frac{\vec{\tau}\cdot\vec{\phi}(x)}{F}\right],
\end{eqnarray}
   where in both cases the three hermitian fields $\pi_i$ and $\phi_i$
describe pion fields transforming as isovectors.
   The connection between the two different choices can be interpreted
as a change of variables, leaving the free-field part of the Lagrangian
unchanged \cite{Chisholm61,Kamefuchi61},
\begin{equation}
\label{ft}
\frac{\vec{\pi}}{F}=\hat{\phi}\sin\left(\frac{\phi}{F}\right)
=\frac{\vec{\phi}}{F}\left(1-\frac{1}{6}\frac{\vec{\phi}^2}{F^2}
+\cdots\right).
\end{equation}

   Let us now consider the tree-level invariant amplitude for
$\pi^+(p_1)+\pi^0(p_2)\to\pi^+(p_3)+\pi^0(p_4)$.
   For that purpose we need to insert the expressions for $U$ of
Eqs.\ (\ref{u1}) and (\ref{u2}) into Eq.\ (\ref{nlsm}) and collect
those terms containing four pion fields:\footnote{The Lagrangian
of Eq.\ (\ref{nlsm}) only generates interaction terms containing
an even number of pion fields, i.e.\ it is even under the substitution
$U\to U^\dagger$ corresponding, respectively, to $\vec{\pi}\to -\vec{\pi}$
and $\vec{\phi}\to -\vec{\phi}$.}
\begin{eqnarray}
{\cal L}_1^{4\pi}&=&\frac{1}{2F^2}\partial_\mu \vec{\pi}\cdot\vec{\pi}
\partial^\mu \vec{\pi}\cdot\vec{\pi}
-\frac{m_\pi^2}{8 F^2}(\vec{\pi}^2)^2,\\
{\cal L}_2^{4\phi}&=&\frac{1}{6F^2}(\partial_\mu \vec{\phi}\cdot\vec{\phi}
\partial^\mu \vec{\phi}\cdot\vec{\phi}-\vec{\phi}^2 \partial_\mu\vec{\phi}
\cdot \partial^\mu\vec{\phi})
+\frac{m_\pi^2}{24 F^2}(\vec{\phi}^2)^2.
\end{eqnarray}
   Observe that the two interaction Lagrangians depend differently on the
respective pion fields.
   Expressing the physical pion fields in terms of the Cartesian components
as $\pi^\pm=\frac{1}{\sqrt{2}}(\pi_1\mp i\pi_2)$ and $\pi^0=\pi_3$ (and
similarly for $\phi_i$), it is straightforward to obtain the corresponding
Feynman rules for $\pi^+(p_1)+\pi^0(p_2)\to\pi^+(p_3)+\pi^0(p_4)$:
\begin{eqnarray}
\label{m1}
{\cal M}_1^{\pi\pi} &=&\frac{i}{F^2}T_0(p_1,p_3),\\
\label{m2}
{\cal M}_2^{\pi\pi} &=&\frac{i}{F^2}\left[T_0(p_1,p_3)-\frac{1}{3}(
\Lambda_1+\Lambda_2+\Lambda_3+\Lambda_4)\right],
\end{eqnarray}
   where $T_0(p_1,p_3)=(p_3-p_1)^2-m_\pi^2$ and $\Lambda_i=p_i^2-m_\pi^2$.
   If initial and final pions are on the mass shell, i.e.\ $\Lambda_i=0$,
the result for the scattering amplitudes is the same which is a consequence
of the equivalence theorem
\cite{Haag58,Chisholm61,Kamefuchi61,Coleman69}.\footnote{For a general
proof of the
equivalence of S-matrix elements evaluated at tree level (phenomenological
approximation), see Sect.\ 2 of Ref.\ \cite{Coleman69}.}
   In fact, since our starting point is the nonlinear $\sigma$ model,
the on-shell result corresponds to the current-algebra prediction
for low-energy $\pi\pi$ scattering \cite{Weinberg67}.
   On the other hand, if one of the momenta of the external lines is off mass
shell, the amplitudes of Eqs.\ (\ref{m1}) and (\ref{m2}) differ.

   According to the standard argument in nucleon-nucleon bremsstrahlung 
one would now
try to discriminate between different on-shell equivalent $\pi\pi$ amplitudes
through an investigation of the reaction
$\pi^+(p_1)+\pi^0(p_2)\to\pi^+(p_3)+\pi^0(p_4)+\gamma(k)$.
   We will now critically examine this claim in the framework of the above
model.
   To that end we include the electromagnetic field as in
Eq.\ (\ref{nlsm}) and calculate the relevant tree-level
diagrams.\footnote{We have
checked that first inserting the parameterizations of $U$ into the
nonlinear $\sigma$ model without electromagnetic interaction
and then performing minimal substitutions $\partial_\mu \pi^\pm\to
(\partial_\mu \pm ie A_\mu)\pi^\pm$ (and similarly for $\phi^\pm$)
generates the same result.}

   Inclusion of the electromagnetic interaction in combination with the first
parameterization of Eq.\ (\ref{u1}) only generates electromagnetic interaction
terms containing two oppositely charged pion fields with either one or two
electromagnetic fields, i.e.\ the interaction is the same as for a point pion
in scalar QED,
\begin{equation}
\frac{F^2}{4}\mbox{Tr}\left[D_\mu U (D^\mu U)^\dagger\right]
=\frac{1}{2}(\partial_\mu \sigma\partial^\mu\sigma
+\partial_\mu\vec{\pi}\cdot\partial^\mu\vec{\pi})
-ieA_\mu(\pi^-\partial^\mu\pi^+-\pi^+\partial^\mu\pi^-)
+e^2 A^2 \pi^+\pi^-,
\end{equation}
   where $\sigma=\sqrt{F^2-\vec{\pi}^2}$.
   This is due to the fact that in Eq.\ (\ref{u1}) the pion field appears
only linearly in combination with the Pauli matrices such that the
commutator with $\tau_3$ in the covariant derivative also results only
in a linear term.
   As a consequence, at tree level only the two diagrams, where the photon
is radiated off the initial and final charged pions, contribute to
the bremsstrahlung amplitude:\footnote{For notational simplicity, we will
omit the complex conjugation of the polarization vectors of final-state
photons.}
\begin{eqnarray}
\label{mbrems1}
{\cal M}_1^{\pi\pi\gamma} &=&\frac{i}{F^2}T_0(p_1-k,p_3)
\frac{i}{(p_1-k)^2-m^2_\pi}
(-2ie p_1\cdot \epsilon)\nonumber
\\&&
-2ie p_3\cdot\epsilon \frac{i}{(p_3+k)^2-m^2_\pi}\frac{i}{F^2}T_0(p_1,
p_3+k)\nonumber\\
&=&
e\left(\frac{p_3\cdot\epsilon}{p_3\cdot k}
-\frac{p_1\cdot \epsilon}{p_1
\cdot k}\right)\frac{i}{F^2}[T_0(p_1,p_3)-2(p_1-p_3)\cdot k].
\end{eqnarray}

   It is now natural to ask how the different off-shell behavior of
the $\pi\pi$ amplitude of Eq.\ (\ref{m2}) enters into the calculation
of the bremsstrahlung amplitude.
   Observe, in this context, that inserting the parameterization
of Eq.\ (\ref{u2}) into Eq.\ (\ref{nlsm}) generates electromagnetic
interactions involving $2n$ pion fields, where n is a positive integer.
   The additional interaction term relevant to the bremsstrahlung process reads
\begin{equation}
\label{l4phiphoton}
{\cal L}_2^{\phi^+\phi^-\phi^0\phi^0 A}=
\frac{ie A_\mu}{3 F^2}(\partial^\mu \phi^+\phi^--\phi^+\partial^\mu \phi^-)
\phi^0\phi^0.
\end{equation}
  Hence the total tree-level amplitude now contains a four-pion-one-photon
contact diagram, in addition to the
two diagrams involving radiation from the charged external legs
which are the same as in ${\cal M}_1^{\pi\pi\gamma}$,
\begin{eqnarray}
\label{mbrems2}
{\cal M}_2^{\pi\pi\gamma} &=&\frac{i}{F^2}\left\{T_0(p_1-k,p_3)
-\frac{1}{3}\left[(p_1-k)^2-m^2_\pi\right]\right\}
\frac{i}{(p_1-k)^2-m_\pi^2}(-2iep_1\cdot\epsilon)\nonumber\\
&&-2iep_3\cdot \epsilon \frac{i}{(p_3+k)^2-m_\pi^2}
\frac{i}{F^2}\left\{T_0(p_1,p_3+k)
-\frac{1}{3}\left[(p_3+k)^2-m_\pi^2\right]\right\}\nonumber\\
&&+\frac{2ie}{3 F^2}(p_1+p_3)\cdot \epsilon.
\end{eqnarray}
    Combining the contribution due to the off-shell behavior in
the $\pi\pi$ amplitude with the contact-term contribution,
we find a precise cancellation of off-shell effects and contact
interactions such that the final result is the same for both
parameterizations, i.e.\ ${\cal M}_1^{\pi\pi\gamma}={\cal M}_2^{\pi\pi\gamma}$.
   This is once again a manifestation of the equivalence theorem
of field theory.
   What is even more important in the present context is the observation
that the two mechanisms, i.e.\ contact term vs.\ off-shell
effects, are indistinguishable since they lead to the same measurable
amplitude.

   It should be noted that none of the above arguments relies on chiral
perturbation theory or is a consequence of chiral symmetry.
   Also, even though gauge invariance poses restrictions on the result
for the bremsstrahlung amplitude it is {\em not} the primary reason for
the equivalence of the two results.
   This will become more evident in the next section, when we include spin
in the problem and consider separately gauge-invariant terms.

\section{Simple Model for the Spin One-Half Case}
\label{sect:half}

We now include spin and consider a very simple model which can describe the
interactions of nucleons and photons and which we can use to describe
nucleon-nucleon bremsstrahlung and Compton scattering on the nucleon.
Such model is described by the Lagrangian
\begin{equation}
{\cal L}_0 = \overline{\Psi}(i D \hspace{-.6em}/ -m) \Psi -\frac{e \kappa}{4 m}
\overline{\Psi} \sigma_{\mu \nu} F^{\mu \nu} \Psi
+\overline{\Phi}(i \partial\hspace{-.5em}/ -m)\Phi
+g \overline{\Psi} \Psi
\overline{\Phi} \Phi.
\label{eq:L0}
\end{equation}
Here $D$ is the covariant derivative $D_\mu \Psi
= (\partial_\mu +ieA_\mu)\Psi$, $e$ and
$\kappa$ are the proton charge and anomalous magnetic moment respectively,
$A_\mu$ is the photon field and $F_{\mu\nu} = \partial_\mu A_\nu - \partial_\nu
A_\mu$ is the electromagnetic field strength tensor.  The fields $\Psi$
correspond to protons and $\Phi$ to neutrons. We separate out the neutrons and
neglect the electromagnetic coupling to the neutron magnetic moment purely to
simplify the calculation of nucleon-nucleon bremsstrahlung, i.e.\ so that we
need consider radiation from only two legs instead of four and so that we do
not need to worry about antisymmetrization for identical particles. This
simplifies the algebra, and reduces the number of diagrams to be considered
from eight to two, but makes no substantive change in what one can learn from
the model.

For the electromagnetic interaction of real photons with protons the above
Lagrangian is exactly what would normally be
used.\footnote{For a more general phenomenological gauge-invariant Lagrangian
capable of describing also the interaction with virtual photons, see Eq.\
(3.1) of Ref.\ \cite{Scherer96}.}
It is gauge invariant, since it involves
only covariant derivatives and field strength tensors. It leads in momentum
space to the standard photon nucleon vertex\footnote{For notational convenience
we will include positive-energy spinors in our expression for vertices
with the understanding that they have to be replaced by appropriate
Feynman propagators where necessary.}
\begin{equation}
-ie \overline{u}(p_f) \left( \slsh{\epsilon} -
\frac{i \kappa}{2m} \sigma_{\mu\nu} \epsilon^\mu k^\nu \right) u(p_i),
\label{eq:pgamvertex}
\end{equation}
corresponding to the momenta $p_i=p_f+k$, i.e.\ for outgoing photons,
and the photon polarization vector $\epsilon$.

For the strong interaction corresponding to a nucleon-nucleon vertex this
Lagrangian gives
\begin{equation}
i g \overline{u}_p(p_3) u_p(p_1) \overline{u}_n(p_4)u_n(p_2),
\label{eq:strongvertex}
\end{equation}
where $p_1$ and $p_3$ correspond to the protons and $p_1+p_2=p_3+p_4$. This
sort of interaction could be generated by the exchange of a heavy scalar
meson. It clearly grossly oversimplifies the strong interaction, but is
sufficient, as we will see, to illustrate all of the principles we wish to
consider.

One should note that this ${\cal L}_0$, at least to lowest order, does not
generate any off-shell effects at either of the strong or electromagnetic
vertices.

Using these vertices we can calculate the Born amplitudes for both
nucleon-nucleon bremsstrahlung and Compton scattering on the nucleon.

We find for nucleon-nucleon bremsstrahlung
\begin{eqnarray}
{\cal M}_0^{NN\gamma}&=&
ieg \overline{u}_n(p_4)u_n(p_2) \overline{u}_p(p_3)\left[ \left(
\slsh{\epsilon} -\frac{i \kappa}{2m} \sigma_{\mu\nu} \epsilon^\mu k^\nu \right)
\frac{\slsh{p_3}+\slsh{k} +m}{(p_3+k)^2 -m^2} \right. \nonumber \\ &+& \left.
\frac{\slsh{p_1}-\slsh{k} +m}{(p_1-k)^2 -m^2}
\left( \slsh{\epsilon} -\frac{i \kappa}{2m} \sigma_{\mu\nu}
\epsilon^\mu k^\nu \right) \right] u_p(p_1).
\label{eq:nngborn}
\end{eqnarray}
This amplitude corresponds to the usual choice for nucleon-nucleon
bremsstrahlung for the electromagnetic parts, but has a much simplified
interaction for the strong part.  Extension to the most general nucleon-nucleon
interaction could be done along the lines of Ref. \cite{hwfnng}, but such
extension adds nothing to the argument here.

Similarly the Born amplitude for Compton scattering on the nucleon, with $p_i +
k_1 = p_f + k_2$, is
\begin{eqnarray}
{\cal M}_0^{CS} &=& -ie^2 \overline{u}(p_f) \left[
\left( \slsh{\epsilon}_2 -\frac{i \kappa}{2m} \sigma_{\mu\nu}
\epsilon^\mu_2 k^\nu_2 \right)
\frac{\slsh{p}_f +\slsh{k}_2 + m}{(p_f+k_2)^2-m^2}
\left( \slsh{\epsilon}_1 +\frac{i \kappa}{2m} \sigma_{\mu\nu}
\epsilon^\mu_1 k^\nu_1 \right) \right. \nonumber \\ &+& \left.
\left( \slsh{\epsilon}_1 +\frac{i \kappa}{2m} \sigma_{\mu\nu}
\epsilon^\mu_1 k^\nu_1 \right)
\frac{\slsh{p}_f -\slsh{k}_1 + m}{(p_f-k_1)^2-m^2}
\left( \slsh{\epsilon}_2 -\frac{i \kappa}{2m} \sigma_{\mu\nu}
\epsilon^\mu_2 k^\nu_2 \right)
\right] u(p_i).
\label{eq:comptborn}
\end{eqnarray}
This amplitude for Compton scattering is exactly what is normally used for the
Born part of the amplitude \cite{Powell49,Lvov93}.

\section{Effect of Field Transformations}
\label{sect:trans}

We now want to consider the effects of a field transformation on the fields
which are contained in the Lagrangian ${\cal L}_0$. Such a transformation
amounts to a change of representation for the fields. It will generate some new
terms in the Lagrangian so that ${\cal L}_0 \rightarrow {\cal L}_0 + \Delta
{\cal L}$. We know from general principles that this change of representation
will not affect any physically measurable results
\cite{Haag58,Chisholm61,Kamefuchi61,Coleman69}. This
means that the physical amplitudes generated from ${\cal L}_0$ and from ${\cal
L}_0 + \Delta {\cal L}$ will be exactly the same, or alternatively that the
sum of all terms containing a contribution obtained from $ \Delta {\cal L}$
will add up to zero.

The general form for $ \Delta {\cal L}$ can be obtained by making the
substitution $\Psi\rightarrow \Psi + \delta \Psi$, where $\delta \Psi$
is not necessarily an infinitesimal transformation. In principle, a
transformation on the neutron is also possible but for simplicity we
discard this possibility since it does not add anything new to our argument.
The resulting Lagrangian becomes
\begin{equation}
{\cal L}_0(\Psi+\delta\Psi) = {\cal L}_0(\Psi) + \Delta{\cal L}(\Psi)
\label{eq:translag}
\end{equation}
with
\begin{eqnarray}
\Delta{\cal L} &=&
\overline{\Psi}\left[ (i D\hspace{-.6em}/ -m) -\frac{e \kappa}{4 m}
\sigma_{\mu \nu} F^{\mu \nu} + g \overline{\Phi} \Phi \right] \delta \Psi
\nonumber \\ &+&
\delta\overline{\Psi}\left[ (i D\hspace{-.6em}/ -m) -\frac{e \kappa}{4 m}
\sigma_{\mu \nu} F^{\mu \nu} + g \overline{\Phi} \Phi \right]  \Psi
\nonumber \\ &+&
\delta\overline{\Psi}\left[ (i D\hspace{-.6em}/ -m) -\frac{e \kappa}{4 m}
\sigma_{\mu \nu} F^{\mu \nu} + g \overline{\Phi} \Phi \right] \delta \Psi.
\label{eq:deltalag}
\end{eqnarray}

Observe that the first term is, up to a total derivative, proportional to the
equation of motion for $\overline{\Psi}$ as obtained from ${\cal L}_0$.  The
second term is proportional to the equation of motion for $\Psi$.  The last
term of Eq.\ (\ref{eq:deltalag}) however is of second order in $\delta\Psi$ and
thus this situation is somewhat more general than that discussed in Ref.\
\cite{Fearing98}. In that case the formalism of ChPT ensured that this last
term was of higher order in the so-called momentum expansion and thus could be
dropped.  Here we have no such expansion criterion and so this term must be
kept.

Now there are two possible approaches. We can take as our Lagrangian ${\cal
L}_0 + \Delta {\cal L}$. In simple cases, e.g.\ the ChPT example discussed in
Ref.\ \cite{Fearing98}, the second order term in Eq.~(\ref{eq:deltalag}) can be
dropped and $\Delta {\cal L}$ is simply proportional to an equation of
motion. In more general situations the second order part would have to be kept
\cite{Scherer95b}. Pieces of $\Delta {\cal L}$ generate off-shell contributions
to the vertex functions and in general other pieces generate contact terms.
However the full contribution of $\Delta {\cal L}$ vanishes and thus there are
no contributions to measurable amplitudes from the complete set of terms from
$\Delta {\cal L}$.  Another way of saying this is that since $\Delta {\cal L}$
originated as a field transformation on ${\cal L}_0$ it can be completely
transformed away, thus eliminating any dependence on both the off-shell pieces
and other pieces involving contact terms.

A second approach, which is more closely allied to phenomenological
calculations is to divide $\Delta {\cal L}$ into two pieces, $\Delta {\cal L} =
\Delta {\cal L}_1 + \Delta {\cal L}_2$ where $\Delta {\cal L}_1$ contains those
pieces which generate off-shell contributions to the various vertices generated
by the Lagrangian, plus perhaps a few contact terms necessary for gauge
invariance, and where $\Delta {\cal L}_2$ contains the remaining terms which
generate pure contact type contributions to the amplitudes for the processes
being considered. Now one can use ${\cal L}_0 + \Delta {\cal L}_1$ as a
phenomenological Lagrangian. It will generate off-shell contributions at the
vertices and in general the amplitude calculated for a physical process will
depend on the coefficients of this part of the Lagrangian, i.e.\ on
coefficients which also multiply the off-shell contributions to the
vertices. This is the procedure adopted in most calculations purporting to
determine sensitivity of amplitudes to off-shell effects.

However by the general result the total contribution of $\Delta {\cal L}$ must
be zero. This means that the Lagrangian ${\cal L}_0 - \Delta {\cal L}_2$ will
give exactly the same measurable amplitudes, depending on the same parameters
of the Lagrangian. However in this case the Lagrangian generates only contact
terms and does not give any off-shell contributions to the various vertices.
Note that the specific way $\Delta {\cal L}$ is split into $\Delta {\cal L}_1$
and $\Delta {\cal L}_2$ will depend on the reaction in question.

Thus we have two Lagrangians, one which gives off-shell contributions to the
vertex functions which are the building blocks for the full amplitude and one
which does not. Both however give exactly the same measurable physical
amplitudes and thus exactly the same dependence of these measured quantities on
the parameters of the Lagrangian. One must thus conclude that the concept of
off-shell contributions to a physical process is just not a meaningful
concept. One can measure coefficients in a particular choice of
phenomenological Lagrangian by comparing with data, but those coefficients
cannot uniquely be associated with the strength off-shell contributions at the
vertices in any meaningful way.

In the next sections we will see explicitly how these principles appear in our
simple model.

\section{Evaluation of the Transformed Lagrangian}
\label{sect:newlag}

Consider in this section a specific transformation or change of representation
of the fields, which has been chosen to generate off-shell contributions at
both strong and electromagnetic vertices in our simple model and to generate a
Lagrangian corresponding to a phenomenological Lagrangian similar to one which
has been used in investigations of off-shell effects.  Thus take
\begin{equation}
\Psi \rightarrow \Psi + \tilde{a}g \overline{\Phi}\Phi\Psi
+\tilde{b}e \sigma_{\mu\nu} F^{\mu\nu} \Psi.
\label{eq:wftrans}
\end{equation}
Here $\tilde{a}$ and $\tilde{b}$ are real constants which determine the overall
strength of the transformations.

This transformation generates $\Delta {\cal L} = \Delta {\cal L}_1 + \Delta
{\cal L}_2$ with
\begin{eqnarray}
\Delta {\cal L}_1 &=& \tilde{a}g \left[ \overline{\Psi}(i
  \stackrel{\leftharpoonup}{\slsh{\partial}} -m -e A \hspace{-.55em}/) \Psi
\overline{\Phi} \Phi + \overline{\Phi} \Phi \overline{\Psi}(i
  \stackrel{\rightharpoonup}{\slsh{\partial}}-m-e A \hspace{-.55em}/)
\Psi \right]
\nonumber \\
&+&\tilde{b}e \left[ \overline{\Psi}(i
  \stackrel{\leftharpoonup}{\slsh{\partial}} -m -eA \hspace{-.55em}/)
\sigma_{\mu\nu} F^{\mu\nu} \Psi + \overline{\Psi}\sigma_{\mu\nu} F^{\mu\nu} 
(i  \stackrel{\rightharpoonup}{\slsh{\partial}}-m-eA \hspace{-.55em}/) 
\Psi \right]
\label{eq:deltaL1}
\end{eqnarray}
and
\begin{eqnarray}
\Delta {\cal L}_2 &=& -eg (\frac{\tilde{a} \kappa}{2m} -2\tilde{b})
\overline{\Phi} \Phi \overline{\Psi}\sigma_{\mu\nu} F^{\mu\nu} \Psi
\nonumber \\
&-&\frac{e^2 \tilde{b} \kappa}{2m}\overline{\Psi}\sigma_{\mu\nu} F^{\mu\nu}
\sigma_{\alpha\beta} F^{\alpha\beta} \Psi \nonumber \\
&+&\tilde{a}\tilde{b}eg \left[ \overline{\Psi}\sigma_{\mu\nu} F^{\mu\nu} (i
  \stackrel{\leftharpoonup}{\slsh{\partial}} -m -eA \hspace{-.55em}/) \Psi
\overline{\Phi} \Phi + \overline{\Phi} \Phi \overline{\Psi}(i
  \stackrel{\rightharpoonup}{\slsh{\partial}}-m-e
A \hspace{-.55em}/)\sigma_{\mu\nu}
F^{\mu\nu} \Psi \right] \nonumber \\
&+& e^2 \tilde{b}^2 \overline{\Psi}\sigma_{\mu\nu} F^{\mu\nu} (i
  \slsh{\partial} -m -e
A \hspace{-.55em}/)\sigma_{\alpha\beta} F^{\alpha\beta} \Psi.
\label{eq:deltaL2}
\end{eqnarray}

Here we have defined $\overline{\Psi} i
\stackrel{\leftharpoonup}{\slsh{\partial}} \equiv -i(\partial_\mu
\overline{\Psi}) \gamma^\mu$.  Both of these contributions to
$\Delta {\cal L}$ have been expressed in terms of
the covariant derivative, which means including some terms proportional to
$A \hspace{-.55em}/$ in $\Delta {\cal L}_1$ rather than
in $\Delta {\cal L}_2$ so that
both pieces will be manifestly gauge invariant. Some terms not contributing to
either nucleon-nucleon bremsstrahlung or Compton scattering have also been
dropped. One would have to keep those terms if one wanted to show, for more
complicated processes than those considered here,
the equivalence of the S-matrix elements obtained from the original
Lagrangian and the transformed Lagrangian.

Consider first the Lagrangian ${\cal L}_0 + \Delta {\cal L}_1$. This generates
some new contributions to the vertices, in addition to those coming from ${\cal
L}_0$ given in Eqs.\ (\ref{eq:pgamvertex}) and (\ref{eq:strongvertex}) above.
For the strong vertex we find from the first term of $ \Delta {\cal L}_1$
\begin{equation}
i \tilde{a}g \overline{u}_p(p_3)[(\slsh{p}_3 -m) + (\slsh{p}_1 -m)] u_p(p_1)
\overline{u}_n(p_4)u_n(p_2).
\label{eq:L1strongvertex}
\end{equation}
Clearly this represents an off-shell contribution to the strong vertex, of
strength determined by the parameter $\tilde{a}$, analogous to what one would
calculate in a potential model for the off-shell nucleon-nucleon vertex. It
vanishes when the momenta $p_1$ and $p_3$ are on shell.

At the electromagnetic vertex we get from the second term in $\Delta {\cal
L}_1$
\begin{equation}
-2 ie \tilde{b}	 \overline{u}(p_f) \left[ (\slsh{p}_f -m)
i  \sigma_{\mu\nu} \epsilon^\mu k^\nu  +
i  \sigma_{\mu\nu} \epsilon^\mu k^\nu (\slsh{p}_i -m) \right] u(p_i).
\label{eq:L1pgamvertex}
\end{equation}
Again this corresponds to an off-shell contribution, this time to the
magnetic part of the photon-nucleon vertex.
The overall strength is determined by the parameter $\tilde{b}$.

Finally there are the contact terms coming from the use of the covariant
derivative as necessary for gauge invariance. The term
\begin{equation}
-2 i \tilde{a}eg \overline{u}_p(p_3) \slsh{\epsilon} u_p(p_1)
\overline{u}_n(p_4)u_n(p_2)
\label{eq:L1acontact}
\end{equation}
corresponds to a photon-four-nucleon vertex and will contribute to
nucleon-nucleon bremsstrahlung. The term
\begin{equation}
+2 ie^2 \tilde{b}  \overline{u}(p_f) \left(
\slsh{\epsilon}_1 i  \sigma_{\mu\nu} \epsilon^\mu_2 k^\nu_2  -
i  \sigma_{\mu\nu} \epsilon^\mu_1 k^\nu_1 \slsh{\epsilon}_2  +
i  \sigma_{\mu\nu} \epsilon^\mu_2 k^\nu_2 \slsh{\epsilon}_1  -
\slsh{\epsilon}_2 i  \sigma_{\mu\nu} \epsilon^\mu_1 k^\nu_1
\right) u(p_i)
\label{eq:L1bcontact}
\end{equation}
gives a two-photon contact vertex which will contribute to Compton
scattering.

This Lagrangian, ${\cal L}_0 + \Delta {\cal L}_1$, corresponds very closely to
some which have been used to investigate off-shell effects in nucleon-nucleon
bremsstrahlung.	 The strong part is simplified, but produces off-shell effects
in the nucleon-nucleon vertex amplitude analogous to those one might obtain
from a potential model. The electromagnetic part is of the general form given
by Bincer \cite{Bincer60} and used by a variety of authors
\cite{Naus87,Herrmann,Doenges,Kondratyuk,Li,Bos93}
to investigate the supposed sensitivity of nucleon-nucleon bremsstrahlung to
off-shell effects. For example in the notation of Nyman \cite{Nyman70},
$F_2^-(m^2) \rightarrow \kappa+8 m^2 \tilde{b}$.\footnote{Because
we work at tree level a distinction between irreducible and reducible
vertex \cite{Naus87} is not necessary}

The alternative Lagrangian ${\cal L}_0 - \Delta {\cal L}_2$ generates just
contact terms. The vertices generated are a contribution to the
one-photon-four-nucleon amplitude from the first and third term
in $\Delta {\cal L}_2$
\begin{eqnarray}
&+&2 ieg \overline{u}_p(p_3) i	\sigma_{\mu\nu} \epsilon^\mu k^\nu
\left( \frac{ \tilde{a} \kappa}{2m}-2\tilde{b} \right) u_p(p_1)
\overline{u}_n(p_4)u_n(p_2) \nonumber \\
&-& 2 i e g \tilde{a} \tilde{b}\overline{u}_p(p_3) \left[(\slsh{p}_1 -
\slsh{k} -m)
i\sigma_{\mu\nu} \epsilon^\mu k^\nu + i\sigma_{\mu\nu} \epsilon^\mu k^\nu
(\slsh{p}_3 + \slsh{k} -m) \right]u_p(p_1)
\overline{u}_n(p_4)u_n(p_2)
\label{eq:L2acontact}
\end{eqnarray}
and a contribution to the two-photon-two-nucleon amplitude from the second and
fourth terms
\begin{eqnarray}
&+& 2i\frac{e^2 \kappa}{m} \tilde{b}\overline{u}(p_f)\left(
i\sigma_{\mu\nu} \epsilon^\mu_1 k^\nu_1
i\sigma_{\alpha\beta} \epsilon^\alpha_2 k^\beta_2
+i\sigma_{\alpha\beta} \epsilon^\alpha_2 k^\beta_2
i\sigma_{\mu\nu} \epsilon^\mu_1 k^\nu_1 \right) u_p(p_i) \nonumber \\
&-& 4 i e^2 \tilde{b}^2	 \overline{u}(p_f) \left[
i\sigma_{\mu\nu} \epsilon^\mu_1 k^\nu_1 (\slsh{p}_i - \slsh{k}_2 -m)
i\sigma_{\alpha\beta} \epsilon^\alpha_2 k^\beta_2 +
i\sigma_{\alpha\beta} \epsilon^\alpha_2 k^\beta_2
(\slsh{p}_i + \slsh{k}_1 -m)i\sigma_{\mu\nu} \epsilon^\mu_1 k^\nu_1 \right]
u(p_i).
\label{eq:L2bcontact}
\end{eqnarray}

Several terms originating in the covariant derivatives have been dropped, as
they contribute to neither bremsstrahlung nor Compton scattering.

   {F}rom the general result that physical amplitudes must be independent of a
field transformation we know that the amplitudes
from ${\cal L}_0$ and ${\cal L}_0
+\Delta{\cal L}$ must be the same.
Thus, to the order we are considering, these two Lagrangians, 
${\cal L}_0 + \Delta {\cal L}_1$ and ${\cal L}_0 - \Delta {\cal L}_2$, 
will give exactly the same physically
measurable amplitudes, yet the first generates off-shell contributions to the
vertices and the second does not. To see this explicitly we must calculate the
amplitudes for these processes in detail, which we will do in the next two
sections.

\section{Explicit Evaluation of Nucleon-Nucleon Bremsstrahlung}
\label{sect:brem}

Consider now the nucleon-nucleon bremsstrahlung process which we will evaluate
using each of the two Lagrangians, ${\cal L}_0 + \Delta {\cal L}_1$ and ${\cal
L}_0 - \Delta {\cal L}_2$. We consider only tree-level diagrams and so must
include radiation from each of the proton legs, with off-shell contributions at
both strong and electromagnetic vertices together with the contact terms
appropriate for each Lagrangian. As noted earlier we treat the $\Phi$ fields as
neutrons and neglect radiation from their magnetic moment. Hence there are only
two diagrams with radiation from external legs.

Consider first the Lagrangian ${\cal L}_0 + \Delta {\cal L}_1$. The
contribution from ${\cal L}_0$ has already been given in
Eq.\ (\ref{eq:nngborn}).
The amplitude coming from the parts of $\Delta {\cal L}_1$ corresponding to
off-shell contributions at strong or electromagnetic vertices, that is the
contribution from ${\cal L}_0$ at one vertex, $\Delta {\cal L}_1$ at the other,
with a propagator in between, is
\begin{equation}
ieg \overline{u}_n(p_4)u_n(p_2) \overline{u}_p(p_3)\left[ 2 \tilde{a}
\slsh{\epsilon} - \left( \frac{\tilde{a} \kappa}{m} -4\tilde{b} \right)
i\sigma_{\mu\nu} \epsilon^\mu k^\nu \right]  u_p(p_1).
\label{eq:nnga}
\end{equation}
The amplitude with off-shell contributions from both strong and electromagnetic
vertices simultaneously is
\begin{equation}
2ieg\tilde{a}\tilde{b} \overline{u}_n(p_4)u_n(p_2) \overline{u}_p(p_3)\left[
i\sigma_{\mu\nu} \epsilon^\mu k^\nu (\slsh{p}_3+\slsh{k} -m) +
(\slsh{p}_1-\slsh{k} -m)
i\sigma_{\mu\nu} \epsilon^\mu k^\nu \right] u_p(p_1).
\label{eq:nngb}
\end{equation}
Finally, the contribution of the contact term originating in the use of
covariant instead of regular derivatives is
\begin{equation}
-2ieg\tilde{a} \overline{u}_n(p_4)u_n(p_2) \overline{u}_p(p_3)\slsh{\epsilon}
 u_p(p_1).
\label{eq:nngc}
\end{equation}
The full amplitude is the sum of Eqs.\ (\ref{eq:nngborn}), (\ref{eq:nnga}),
(\ref{eq:nngb}) and (\ref{eq:nngc}).  Note that the contact term of
Eq. (\ref{eq:nngc}) actually cancels the similar term coming from the off-shell
contributions at the strong vertex in Eq.\ (\ref{eq:nnga}) and that the net
result coming from $\Delta {\cal L}_1$ consists of a number of magnetic type
terms proportional to $\sigma_{\mu\nu} \epsilon^\mu k^\nu$.

   {F}rom one viewpoint this Lagrangian, ${\cal L}_0 + \Delta {\cal L}_1$, can
be considered simply as a purely phenomenological Lagrangian. It leads to a
bremsstrahlung amplitude, which can be compared with experiment so as to
extract values of the phenomenological parameters $\tilde{a}, \tilde{b}$. Such
an approach is perfectly acceptable as long as one is completely clear that the
Lagrangian is just phenomenological. Its usefulness will depend on how close
the model Lagrangian reproduces the real physical situation. Difficulties arise
however when one makes the traditional, though as we shall see incorrect, claim
that the values of $\tilde{a}, \tilde{b}$ so obtained correspond to some
measure of off-shell effects.

To see how this claim arises and why it is incorrect let us first see how our
simple model is closely analogous to the traditional approaches. Thus for
example in standard nonrelativistic potential model approaches to
nucleon-nucleon bremsstrahlung, one would first calculate in the abstract an
off-shell nucleon-nucleon amplitude corresponding to a potential.  This would
give a result analogous to the off-shell amplitude of Eq.\
(\ref{eq:L1strongvertex}) calculated in our model. Different
on-shell-equivalent potentials could still give different amplitudes,
corresponding to different values of $\tilde{a}$.

Similarly at the electromagnetic vertex one traditionally parameterizes the
off-shell behavior in some way, or uses some model to obtain something
analogous to Eq. (\ref{eq:L1pgamvertex}). Various authors have used dispersion
relations, simple pion-loop models, chiral perturbation theory, or purely
phenomenological considerations. In all cases the abstract off-shell
electromagnetic vertex is governed by a strength parameter similar to our
$\tilde{b}$.

Thus in these traditional approaches one argues that in the abstract the
off-shell contributions to the strong and electromagnetic interaction vertices
are proportional to $\tilde{a}$ and $\tilde{b}$ respectively.  These are
parameters of the Lagrangian, which appear in the amplitude, and can thus be
determined from a comparison of the amplitude with measured
quantities. Therefore, one traditionally concludes the values of these
parameters measure in some physical way off-shell behavior in the strong and
electromagnetic interactions.

It is this last part of the argument which is incorrect. In actual fact the
values of $\tilde{a}$ and $\tilde{b}$ tell us nothing, in any unambiguous way,
about off-shell behavior. To see this in detail let us calculate the
bremsstrahlung amplitude using the alternative Lagrangian ${\cal L}_0 - \Delta
{\cal L}_2$.  We obtain from $-\Delta {\cal L}_2$ two contact terms. {F}rom the
first term of Eq.\ (\ref{eq:L2acontact})
\begin{equation}
-ieg \left( \frac{\tilde{a} \kappa}{m} -4\tilde{b} \right)
\overline{u}_n(p_4)u_n(p_2) \overline{u}_p(p_3)
i\sigma_{\mu\nu} \epsilon^\mu k^\nu   u_p(p_1)
\label{eq:nngcont1}
\end{equation}
and from the second
\begin{equation}
2ieg\tilde{a}\tilde{b} \overline{u}_n(p_4)u_n(p_2) \overline{u}_p(p_3)\left[
i\sigma_{\mu\nu} \epsilon^\mu k^\nu (\slsh{p}_3+\slsh{k} -m) +
(\slsh{p}_1-\slsh{k} -m)
i\sigma_{\mu\nu} \epsilon^\mu k^\nu \right] u_p(p_1).
\label{eq:nngcont2}
\end{equation}

Clearly the full amplitude from ${\cal L}_0 - \Delta {\cal L}_2$,
Eqs.\ (\ref{eq:nngborn}), (\ref{eq:nngcont1}) and (\ref{eq:nngcont2}) is
exactly the same as that from ${\cal L}_0 + \Delta {\cal L}_1$,
Eqs.\ (\ref{eq:nngborn}), (\ref{eq:nnga}), (\ref{eq:nngb}) and (\ref{eq:nngc}),
as it must be, by virtue of the general results from field transformation
arguments. Thus a comparison with data using this Lagrangian will produce
exactly the same values of $\tilde{a}, \tilde{b}$ as with the original
Lagrangian. However ${\cal L}_0 - \Delta {\cal L}_2$ produces no off-shell
effects at either strong or electromagnetic vertices, so there cannot be any
meaningful connection between the values of $\tilde{a}, \tilde{b}$ and
off-shell effects.

More generally consider a combination of these two Lagrangians, ${\cal L}(\eta)
= {\cal L}_0 +(1-\eta)\Delta{\cal L}_1 - \eta \Delta {\cal L}_2$ where $\eta$
is an arbitrary real parameter.
Since $\Delta{\cal L}_1$
and $- \Delta {\cal L}_2$ produce the same amplitudes, the result will be
independent of $\eta$.	Thus ${\cal L}(\eta)$ for arbitrary $\eta$ will all
lead to the same bremsstrahlung amplitude, that given by
Eqs.\ (\ref{eq:nngborn}), (\ref{eq:nnga}), (\ref{eq:nngb}) and (\ref{eq:nngc}),
and hence to exactly the same values of $\tilde{a}, \tilde{b}$ if this
amplitude is compared to data. However the off-shell strong and electromagnetic
vertices calculated with ${\cal L}(\eta)$ will have strengths
$(1-\eta)\tilde{a}$ and $(1-\eta)\tilde{b}$ respectively. Thus a single set of
$\tilde{a}, \tilde{b}$ corresponds to arbitrary values of the off-shell
vertices. Thus it should be very clear that such off-shell behavior is not a
physically measurable quantity. It makes no sense to equate sensitivity to
$\tilde{a}, \tilde{b}$ with sensitivity to off-shell behavior, or to claim to
be able to measure off-shell behavior by measuring parameters appearing in the
amplitude, as has traditionally been done in discussions of off-shell behavior.

In retrospect this result perhaps should not be surprising. The concept of an
off-shell amplitude is in some sense a mathematical concept, which applies only
to a piece of a diagram. An off-shell particle is not physical and one can
never measure it directly. Such amplitudes have meaning only when put into a
larger diagram which has appropriate interactions to put the particle back on
shell. Thus one should perhaps expect a large measure of ambiguity in
describing such an intermediate, unphysical state.

\section{Explicit Evaluation of Compton Scattering}
\label{sect:compton}

The Lagrangian which has just been used to evaluate nucleon-nucleon
bremsstrahlung also leads to an amplitude for Compton scattering. Only the
electromagnetic part is required, so the simplification of the strong
interaction used for nucleon-nucleon bremsstrahlung is not necessary.  In this
section we evaluate that amplitude explicitly and will find a situation exactly
similar to that of nucleon-nucleon bremsstrahlung. Namely, since the total
contribution from the part of the Lagrangian generated by the field
transformation, $\Delta {\cal L}_1 + \Delta {\cal L}_2$, vanishes there are two
alternate Lagrangians (actually an infinite set of linear combinations) which
give the same Compton amplitude, ${\cal L}_0 + \Delta {\cal L}_1$ which
produces off-shell contributions in the electromagnetic vertices and 
${\cal L}_0 - \Delta {\cal L}_2$ which produces only contact terms.

Consider first the Compton amplitude originating in the Lagrangian ${\cal L}_0
+ \Delta {\cal L}_1$. This Lagrangian is essentially identical to that used in
phenomenological calculations where a phenomenological off-shell part is
introduced at the electromagnetic gamma-nucleon-nucleon vertex. The
contribution from ${\cal L}_0$ has been given in Eq.\ (\ref{eq:comptborn}).
Just as for nucleon-nucleon bremsstrahlung the contribution from
$\Delta {\cal L}_1$
leads to an amplitude with an off-shell contribution at one or the other of the
electromagnetic vertices of the form
\begin{eqnarray}
 &+& 2 i\tilde{b} e^2 \overline{u}(p_f) \left[
\left( \slsh{\epsilon}_2 -\frac{i \kappa}{2m} \sigma_{\mu\nu}
\epsilon^\mu_2 k^\nu_2 \right)	i\sigma_{\alpha\beta}
\epsilon^\alpha_1 k^\beta_1 -i \sigma_{\alpha\beta}
\epsilon^\alpha_2 k^\beta_2 \left( \slsh{\epsilon}_1 +\frac{i \kappa}{2m}
\sigma_{\mu\nu}
\epsilon^\mu_1 k^\nu_1 \right) \right] u(p_i) \nonumber \\
 &-& 2 i\tilde{b} e^2 \overline{u}(p_f) \left[
\left( \slsh{\epsilon}_1 +\frac{i \kappa}{2m} \sigma_{\mu\nu}
\epsilon^\mu_1 k^\nu_1 \right)	i\sigma_{\alpha\beta}
\epsilon^\alpha_2 k^\beta_2 -i \sigma_{\alpha\beta}
\epsilon^\alpha_1 k^\beta_1 \left( \slsh{\epsilon}_2 -\frac{i \kappa}{2m}
\sigma_{\mu\nu}
\epsilon^\mu_2 k^\nu_2 \right) \right] u(p_i)
\label{eq:compta}
\end{eqnarray}
and a contribution with an off-shell part of the vertex at both electromagnetic
vertices given by
\begin{eqnarray}
&+& 4 i \tilde{b}^2 e^2 \overline{u}(p_f) \left[i \sigma_{\mu\nu}
\epsilon^\mu_2 k^\nu_2 (\slsh{p}_f+\slsh{k}_2 -m)i \sigma_{\alpha\beta}
\epsilon^\alpha_1 k^\beta_1 \right] u(p_i)  \nonumber \\
&+& 4 i \tilde{b}^2 e^2 \overline{u}(p_f) \left[i \sigma_{\mu\nu}
\epsilon^\mu_1 k^\nu_1 (\slsh{p}_f-\slsh{k}_1 -m)i \sigma_{\alpha\beta}
\epsilon^\alpha_2 k^\beta_2 \right] u(p_i).
\label{eq:comptb}
\end{eqnarray}
There is also a two-photon contact term coming from the use of the covariant
derivative in the last part of $\Delta {\cal L}_1$ which can be written as
\begin{equation}
+2 ie^2 \tilde{b}  \overline{u}(p_f) \left(
\slsh{\epsilon}_1 i  \sigma_{\mu\nu} \epsilon^\mu_2 k^\nu_2  -
i  \sigma_{\mu\nu} \epsilon^\mu_1 k^\nu_1 \slsh{\epsilon}_2  +
i  \sigma_{\mu\nu} \epsilon^\mu_2 k^\nu_2 \slsh{\epsilon}_1  -
\slsh{\epsilon}_2 i  \sigma_{\mu\nu} \epsilon^\mu_1 k^\nu_1
\right) u(p_i).
\label{eq:comptc}
\end{equation}
Note that, just as in nucleon-nucleon bremsstrahlung, this contact term cancels
the $\slsh{\epsilon}$ terms from the off-shell contribution of
Eq.\ (\ref{eq:compta}) above, leaving purely magnetic contributions.

The full Compton amplitude originating in ${\cal L}_0 + \Delta {\cal L}_1$ is
then given by the sum of Eqs.\ (\ref{eq:comptborn}), (\ref{eq:compta}),
(\ref{eq:comptb}), and (\ref{eq:comptc}). This corresponds exactly to the
Lagrangian which has been used in phenomenological calculations. Naively in
such an approach, one notes the appearance of the parameter $\tilde{b}$ and
also the fact that it appears in the off-shell electromagnetic vertex and thus
one might, as in nucleon-nucleon bremsstrahlung,  claim to be able to
'determine' the parameter $\tilde{b}$ and thus the off-shell vertex
by a measurement of Compton scattering.

However, just as in the nucleon-nucleon bremsstrahlung case discussed above,
this has to be wrong. We have the equivalent Lagrangian ${\cal L}_0 - \Delta
{\cal L}_2$ which involves only contact terms, but gives the same measurable
result. We can see this specifically. The contributions from $-\Delta {\cal
L}_2$ to the Compton amplitude are just those of Eq.\ (\ref{eq:L2bcontact}),
i.e.
\begin{eqnarray}
&-& 2i\frac{e^2 \kappa}{m} \tilde{b}\overline{u}(p_f)\left(
i\sigma_{\mu\nu} \epsilon^\mu_1 k^\nu_1
i\sigma_{\alpha\beta} \epsilon^\alpha_2 k^\beta_2
+i\sigma_{\alpha\beta} \epsilon^\alpha_2 k^\beta_2
i\sigma_{\mu\nu} \epsilon^\mu_1 k^\nu_1 \right) u_p(p_i) \nonumber \\
&+& 4 i e^2 \tilde{b}^2	 \overline{u}(p_f) \left[
i\sigma_{\mu\nu} \epsilon^\mu_1 k^\nu_1 (\slsh{p}_i - \slsh{k}_2 -m)
i\sigma_{\alpha\beta} \epsilon^\alpha_2 k^\beta_2 +
i\sigma_{\alpha\beta} \epsilon^\alpha_2 k^\beta_2
(\slsh{p}_i + \slsh{k}_1 -m)i\sigma_{\mu\nu} \epsilon^\mu_1 k^\nu_1 \right]
u(p_i).
\label{eq:comptcont}
\end{eqnarray}

Clearly this is the same amplitude as generated by $\Delta {\cal L}_1$.	 Now
the argument is exactly the same as given at the end of the previous
section. The same measurable amplitude is produced by a Lagrangian which
generates an off-shell component at the photon-nucleon-nucleon vertex as by the
one which has only contact terms, or in fact by any linear combination of the
two. Thus the constant $\tilde{b}$ appearing in the Lagrangian is not in any
way a 'measure' of off-shell behavior.

There is another interesting observation one can make using this simple model,
though one somewhat peripheral to the main line of argument. Consider the
Lagrangian ${\cal L}_0 + \Delta {\cal L}_1$ which corresponds to a standard
phenomenological Lagrangian used to describe the photon-nucleon-nucleon vertex.
Sometimes the argument is given that even if the coefficient $\tilde{b}$ does
not represent an off-shell effect it can be used to parameterize the unknown
features of the interaction. That is certainly possible. However such
parameterizations have implications for a variety of processes described by the
Lagrangian and do not always lead to results consistent with the data.

For example with this Lagrangian we get the full amplitude for Compton
scattering from the sum of Eqs.\ (\ref{eq:comptborn}), (\ref{eq:compta}),
(\ref{eq:comptb}) and (\ref{eq:comptc}) and can extract from that amplitude an
expression for the proton electromagnetic polarizabilities, for which there is
some experimental data. To do this we make a two-component reduction of this
amplitude in the lab frame using Coulomb gauge which leads to
\begin{eqnarray}
{\cal M}^{CS}
&=& i \chi^\dagger_f \left[ -\frac{e^2}{m} \vec{\epsilon}_1 \cdot
\vec{\epsilon}_2 + \cdots \right. \\
& & \left. + \omega_1\omega_2\vec{\epsilon}_1 \cdot
\vec{\epsilon}_2 \left( -\frac{4 e^2}{m} (\kappa
\tilde{b} + 4 m^2 \tilde{b}^2) \right) + \omega_1\omega_2\vec{\epsilon}_2
\times \hat{k}_2 \cdot \vec{\epsilon}_1 \times \hat{k}_1
\left( \frac{4 e^2 \kappa \tilde{b}}{m}
\right) \right] \chi,
\label{eq:comptnonrel}
\end{eqnarray}
where the $\chi$'s are two-component nucleon spinors and where we have kept
only the leading Thomson term and those terms contributing to the
polarizabilities (for details see, e.g., Eqs.\ (4) and (6) of Ref.\
\cite{Lvov93}).

{F}rom this we can extract the electromagnetic polarizabilities
\cite{Lvov93} as
\begin{eqnarray}
\overline{\alpha} &=&  -\frac{ e^2}{4 \pi} \frac{4}{m} (\kappa
\tilde{b} + 4 m^2 \tilde{b}^2), \\
\overline{\beta} &=& \frac{e^2}{4 \pi} \frac{4 \kappa \tilde{b}}{m},
\label{eq:polariz}
\end{eqnarray}
where the factor of $4 \pi$ has been inserted to put the polarizabilities into
conventional units, since in our notation $e^2/4 \pi \approx 1/137$.

   Now using the experimental value for the proton polarizability,
$\overline{\alpha} = (12.1\pm 0.8\pm 0.5)\times 10^{-4}\,\mbox{fm}^3$
from Ref.\ \cite{MacGibbon95} we try to solve for
$\tilde{b}$. We find
\begin{equation}
\tilde{b} = \frac{\kappa}{8 m^2}\left( -1 \pm \sqrt{1-\frac{16 \pi m^3
\overline{\alpha}}{e^2 \kappa^2}} \right).
\label{eq:bsoln}
\end{equation}

Since $16 \pi  m^3 \overline{\alpha}/e^2 \kappa^2 = 22 >> 1$ there is
no real solution possible for $\tilde{b}$. This means that simply
introducing a phenomenological term in the Lagrangian which produces an
off-shell photon-nucleon-nucleon vertex, as has been done in many
calculations can lead to results which are inconsistent with data for related
reactions. Clearly even at the phenomenological level one must be consistent
and check the prediction of such phenomenological terms with other processes
generated by the same Lagrangian.

\section{Summary and Conclusions}
\label{sect:summary}
In this paper we have used the concept of field transformations and several
simple models to illustrate the impossibility of measuring off-shell effects in
nucleon-nucleon bremsstrahlung, Compton scattering, and by implication other
medium-energy processes. In the first example, the non-linear $\sigma$ model
Lagrangian, together with two standard representations of the pion field, was
used for pion-pion bremsstrahlung. We showed by specific example that these two
representations gave different off-shell scattering amplitudes, but exactly the
same measurable amplitude for the bremsstrahlung. Thus the measurable quantity
clearly cannot distinguish different off-shell behavior at the vertices making
up the full process.

For the rest of the paper we looked at a spin one-half model Lagrangian. Again
as a result of a field transformation we could generate a modified Lagrangian
which produced an off-shell contribution to both strong and electromagnetic
vertices. This Lagrangian was very closely related to the kinds of
phenomenological Lagrangians which have been used to investigate off-shell
effects at these vertices. Again we showed by specific example that changes in
the Lagrangian which produce different off-shell vertices lead to exactly the
same measurable amplitude. In effect we can interpret pieces of the measurable
amplitude as off-shell effects at the vertices or as contact terms or as any
combination of the two. Thus the concept of 'off-shell vertex', while perfectly
well defined as a mathematical abstraction, does not translate into a
physically measurable quantity. In short, off-shell effects are not measurable.
The model used here properly included spin one-half, did not depend on ChPT, or
on a momentum expansion, and did not depend on gauge invariance, which should
put to rest any possible concerns that the original considerations
of Refs.\ \cite{Scherer95a,Fearing98} depended somehow on the simplicity of the
model considered there.

Finally we should comment that it is certainly possible to construct a
microscopic model of a process which will generate an off-shell form factor at
the vertices, and perhaps some contact terms.  This is in fact standard
procedure. It is a legitimate approach and we can speak of the off-shell form
factor in this model, just as we speak of the off-shell amplitude generated
from a potential by solving the Lippman-Schwinger equation. The generation of
such a off-shell form factor is unique and a function only of the properties of
the model. However one cannot measure this off-shell form factor and thus
cannot determine the correctness or incorrectness of the model based on its
prediction for the off-shell form factor.

This work was supported in part by a grant from the Natural Sciences and
Engineering Research Council of Canada and by the Deutsche
Forschungsgemeinschaft (SFB 443).

\end{document}